# Polynomial-time Computation
## via
# Local Inference Relations

**Robert Givan** and **David McAllester**

**Robert Givan**
Electrical & Computer Engineering
Purdue University
1285 EE Building
West Lafayette, IN 47907
Phone: (765)494-9068
Email: givan@ecn.purdue.edu
Web: http://www.ece.purdue.edu/~givan/

**David McAllester**
P. O. Box 971
AT&T Labs Research
180 Park Avenue
Florham Park, NJ, 07932 USA
Phone: (973)-360-8318
Email: dmac@research.att.com
Web: http://www.research.att.com/~dmac/

**Abstract**

We consider the concept of a *local* set of inference rules. A local rule set can be automatically transformed into a rule set for which bottom-up evaluation terminates in polynomial time. The local-rule-set transformation gives polynomial-time evaluation strategies for a large variety of rule sets that cannot be given terminating evaluation strategies by any other known automatic technique. This paper discusses three new results. First, it is shown that every polynomial-time predicate can be defined by an (unstratified) local rule set. Second, a new machine-recognizable subclass of the local rule sets is identified. Finally we show that locality, as a property of rule sets, is undecidable in general.

**Keywords:** Descriptive Complexity Theory, Decision Procedures, Automated Reasoning,

## 1. Introduction

Under what conditions does a given set of inference rules define a computationally tractable inference relation? This is a syntactic question about syntactic inference rules. There are a variety of motivations for identifying tractable inference relations. First, tractable inference relations sometimes provide decision procedures for semantic theories. For example, the equational inference rules of reflexivity, symmetry, transitivity, and substitutivity define a tractable inference relation that yields a decision procedure for the entailment relation between sets of ground equations [Kozen, 1977], [Shostak, 1978]. Another example is the set of equational Horn clauses valid in lattice theory. As a special case of the results in this paper one can show automatically that validity of a lattice-theoretic Horn clause is decidable in cubic time.

Deductive databases provide a second motivation for studying tractable inference relations. A deductive database is designed to answer queries using simple inference rules as well as a set of declared data base facts. The inference rules in a deductive database typically define a tractable inference relation—these inference rules are usually of a special form known as a datalog program. A datalog program is a set of first-order Horn clauses that do not contain function symbols. Any data-log program defines a tractable inference relation [Ullman, 1988], [Ullman, 1989]. There has been interest in generalizing the inference rules used in deductive databases beyond the special case of datalog programs. In the general case, where function symbols are allowed in Horn clause inference rules, a set of inference rules can be viewed as a Prolog program. Considerable work has been done on "bottom-up" evaluation strategies for these programs and source-to-source transformations that make such bottom-up evaluation strategies more efficient [Naughton and Ramakrishnan, 1991], [Bry, 1990]. The work presented here on local inference relations can be viewed as an extension of these optimization techniques. For example, locality testing provides an automatic source-to-source transformation on the inference rules for equality (symmetry, reflexivity, transitive, and substitution) that allows them to be completely evaluated in a bottom-up fashion in cubic time. We do not know of any other automatic transformation on inference rules that provides a terminating evaluation strategy for this rule set.

Tractable rule sets also play an important role in type-inference systems for computer programming languages [Talpin and Jouvelot, 1992], [Jouvelot and Gifford, 1991]. Although we have not yet investigated connections between the notion of locality used here and known results on tractability for type inference systems, this seems like a fruitful area for future research. From a practical perspective it seems possible that general-purpose bottom-up evaluation strategies for inference rules can be applied to inference rules for type-inference systems. From a theoretical perspective we show below that any polynomial-time predicate can be defined by a local set of inference rules and that many type-inference systems give polynomial-time decidable typability.

A fourth motivation for the study of tractable inference relations is the role that such relations can play in improving the efficiency of search. Many practical search algorithms use some form of incomplete inference to prune nodes in the search tree [Knuth, 1975], [Mackworth, 1977], [Pearl and Korf, 1987]. Incomplete inference also plays an important role in pruning search in constraint logic programming [Jaffar and Lassez, 1987], [van Hentenryck, 1989], [McAllester and Siskind, 1991]. Tractable inference relations can also be used to define a notion of "obvious inference" which can then be used in "Socratic" proof verification systems which require proofs to be reduced to obvious steps [McAllester, 1989], [Givan et al., 1991].

As mentioned above, inference rules are syntactically similar to first-order Horn clauses. In fact, most inference rules can be syntactically represented by a Horn clause in sorted first-order logic. If $R$ is a set of Horn clauses, $\Sigma$ is a set of ground atomic formulas, and $\Phi$ is a ground atomic formula, then we write $\Sigma \vdash_R \Phi$ if $\Sigma \cup R \vdash \Phi$ in first order logic. We write $\vdash_R$ rather than $\models_R$ because we think of $R$ as a set of syntactic inference rules and $\vdash_R$ as the inference relation generated by those rules. Throughout this paper we use the term "rule set" as a synonym for "finite set of Horn clauses". We give nontrivial conditions on $R$ which ensure that the inference relation $\vdash_R$ is polynomial-time decidable.

As noted above, a rule set *R* that does not contain any function symbols is called a datalog program. It is well-known that the inference relation defined by a datalog program is polynomial-time decidable. Vardi and Immerman independently proved, in essence, that datalog programs provide a characterization of the complexity class *P* — any polynomial time predicate on finite databases can be written as a datalog program provided that one is given a successor relation that defines a total order on the domain elements [Vardi, 1982], [Immerman, 1986], [Papadimitriou, 1985].

Although datalog programs provide an interesting class of polynomial-time inference relations, the class of tractable rule sets is much larger than the class of datalog programs. First of all, one can generalize the concept of a datalog program to the concept of a *superficial* rule set. We call a set of Horn clauses superficial if any term that appears in the conclusion of a clause also appears in some antecedent of that clause. A superficial rule set has the property that forward-chaining inference does not introduce new terms. We show in this paper that superficial rule sets provide a different characterization of the complexity class *P*. While datalog programs can encode any polynomial-time predicate on finite databases, superficial rule sets can encode any polynomial-time predicate on first-order terms. Let $\tilde{Q}$ be a predicate on first-order terms constructed from a finite signature. We define the DAG size of a first-order term *t* to be the number of distinct terms that appear as subexpressions of *t*.[1] It is possible to show that if $\tilde{Q}$ can be computed in polynomial time in the sum of the DAG size of its arguments then $\tilde{Q}$ can be represented by a superficial rule set. More specifically, we prove below that for any such predicate $\tilde{Q}$ on *k* first-order terms there exists a superficial rule set *R* such that $\tilde{Q}(t_1, \ldots, t_k)$ if and only if `Input`$(t_1, \ldots, t_k) \vdash_R$ `Accept` where `Input` is a predicate symbol and `Accept` is a distinguished proposition symbol. Our characterization of the complexity class *P* in terms of superficial rule sets differs from Immerman's characterization of *P* in terms of datalog programs in two ways. First, the result is stated in terms of predicates on terms rather than predicates on databases. Second, unlike the datalog characterization, no separate total order on domain elements is required.

Superficial rule sets are a special case of the more general class of *local* rule sets [McAllester, 1993]. A set *R* of Horn clauses is local if whenever $\Sigma \vdash_R \Phi$ there exists a proof of $\Phi$ from $\Sigma$ such that every term in the proof is mentioned in $\Sigma$ or $\Phi$. If *R* is local then $\vdash_R$ is polynomial-time decidable. All superficial rule sets are local but many local rule sets are not superficial. The set of the four inference rules for equality is local but not superficial. The local inference relations provide a third characterization of the complexity class *P*. Let $\tilde{Q}$ be a predicate on first-order terms constructed from a finite signature. If $\tilde{Q}$ can be computed in polynomial time in the sum of the DAG size of its arguments then there exists a local rule set $R$ such that for any terms $t_1$, ..., $t_k$ we have that $Q(t_1, ..., t_k)$ if and only if $\vdash_R Q(t_1, ..., t_k)$ where $Q$ is a predicate symbol representing $\tilde{Q}$. Note that no superficial rule set can have this property because forward-chaining inference from a superficial rule set can not introduce new terms. We find the characterization of polynomial time predicates in terms of local rule sets to be particularly pleasing because as just described it yields a direct mapping from semantic predicates to predicates used in the inference rules.

---

1. The DAG size of a term is the size of the Directed Acyclic Graph representation of the term.

Unlike superficiality, locality can be difficult to recognize. The set of four inference rules for equality is local but the proof of this fact is nontrivial. Mechanically-recognizable subclasses of local rule sets have been identified by McAllester [McAllester, 1993] and Basin and Ganzinger [Basin and Ganzinger, 2000]. Here we introduce a third mechanically-recognizable subclass which contains a variety of natural rule sets not contained in either of these earlier classes. We will briefly describe the two earlier classes and give examples of rules sets included in our new class that are not included in the earlier classes.

Basin and Ganzinger identify the class of rule sets that are saturated with respect to all orderings compatible with the subterm ordering. The notion of saturation is derived from ordered resolution. We will refer to these rule sets simply as "saturated". Saturation with respect to the class of subterm-compatible orders turns out to be a decidable property of rule sets. Membership in the [McAllester, 1993] class or the new class identified here is only semi-decidable --- a rule set is in these classes if there exists a proof of locality of a certain restricted form (a different form for each of the two classes). Unfortunately, although saturation is decidable, it appears that most natural local rule sets are not saturated. Typically they must be expanded (saturated) to include derived rules. This saturation process typically yields a less efficient decision procedure for the inference relation defined by the rules --- the decision procedure must use more inference rules in the larger saturated set. As an example consider the following rules.

$$
\begin{aligned}
& x \leq y, y \leq z \Rightarrow x \leq z \\
& x \leq y \Rightarrow f(x) \leq f(y)
\end{aligned}
\qquad (1)
$$

These rules are local and this rule set is in both McAllester's class and the new class introduced here. But they are not saturated. Saturation adds (at least) the following rules.

$$
\begin{aligned}
& x \leq z, y \leq f(x) \Rightarrow y \leq f(z) \\
& z \leq x, f(x) \leq y \Rightarrow f(z) \leq y
\end{aligned}
\qquad (2)
$$

A decision procedure based on the larger saturated set would still run in $O(n^3)$ time, but the added rules significantly impact the constant factors and this is an important issue in practice.

The mechanically-recognizable subclass of local rule sets introduced in [McAllester, 1993] is called the bounded-local rule sets. This subclass is defined carefully in the body of this paper for further comparison to the new subclass introduced here. The set of the four basic rules for equality is bounded-local. As another example of a bounded-local rule set we give the following rules for reasoning about a monotone operator from sets to sets. Let $R_f$ be the following set of inference rules for a monotone operator.

$$\begin{aligned} & x \leq x \\ & x \leq y, y \leq z \Rightarrow x \leq z \\ & x \leq y \Rightarrow f(x) \leq f(y) \end{aligned} \quad (3)$$

There is a simple source-to-source transformation on any local rule set that converts the rule set to a superficial rule set without changing the relation described. For example, consider the above rules for a monotone operator. We can transform these rules so that they can only derive information about terms explicitly mentioned in the query. To do this we introduce another predicate symbol M (with the intuitive meaning "mentioned"). Let $R'_f$ be the following transformed version of $R_f$.

$$\begin{aligned} & \text{M}(f(x)) \Rightarrow \text{M}(x) \\ & x \leq y \Rightarrow \text{M}(x) \\ & x \leq y \Rightarrow \text{M}(y) \\ & \text{M}(x) \Rightarrow x \leq x \\ & \text{M}(x), \text{M}(y), \text{M}(z), x \leq y, y \leq z \Rightarrow x \leq z \\ & \text{M}(f(x)), \text{M}(f(y)), x \leq y \Rightarrow f(x) \leq f(y) \end{aligned} \quad (4)$$

Note that $R'_f$ is superficial and hence bottom-up (forward-chaining) evaluation must terminate in polynomial time[2]. Then to determine if $\Sigma \vdash_{R_f} t \leq u$ we determine, by bottom-up evaluation whether $\{\text{M}(t), \text{M}(u)\} \cup \Sigma \vdash_{R'_f} t \leq u$. An analogous transformation applies to any local rule set.

A variety of other bounded-local rule sets are given [McAllester, 1993]. As an example of a rule set that is local but not bounded local we give the following rules for reasoning about a lattice.

---

2. For this rule set bottom-up evaluation can be run to completion in cubic time.

$$x \leq x$$
$$x \leq y, y \leq z \Rightarrow x \leq z$$
$$x \leq x \vee y$$
$$y \leq x \vee y \tag{5}$$
$$x \leq z, y \leq z \Rightarrow x \vee y \leq z$$
$$x \wedge y \leq x$$
$$x \wedge y \leq y$$
$$z \leq x, z \leq y \Rightarrow z \leq x \wedge y$$

These rules remain local when the above monotonicity rule is added. With or without the monotonicity rule, the rule set is not bounded-local.

In this paper we construct another machine-recognizable subclass of the local rule sets which we call *inductively-local* rule sets. All of the bounded-local rule sets given in [McAllester, 1993] are also inductively-local. The procedure for recognizing inductively-local rule sets has been implemented and has been used to determine that the above rule set is inductively-local. Hence the inference relation defined by the rules in (5) is polynomial-time decidable. Since these rules are complete for lattices this result implies that validity for lattice-theoretic Horn clauses is polynomial-time decidable.

We been able to show that there are bounded-local rule sets which are not inductively-local, although our examples are somewhat artificial. We have not found any natural examples of local rule sets that fail to be inductively-local. Inductively local rule sets provide a variety of mechanically recognizable polynomial time inference relations.

In this paper we also settle an open question from the previous analysis in [McAllester, 1993] and show that locality as a general property of rule sets is undecidable. Hence the optimization of logic programs based on the recognition of locality is necessarily a somewhat heuristic process.

## 2. BASIC TERMINOLOGY

In this section we give more precise definitions of the concepts discussed in the introduction.

**Definition 1.** A Horn clause is a first order formula of the form $\Psi_1 \wedge \ldots \wedge \Psi_n \Rightarrow \Phi$ where $\Phi$ and the $\Psi_i$ are atomic formulas. For any set of Horn clauses $R$, any finite set $\Sigma$ of ground terms, and any ground atomic formula $\Phi$, we write $\Sigma \vdash_R \Phi$ whenever $\Sigma \cup U(R) \vdash \Phi$ in first-order logic where $U(R)$ is the set of universal closures of Horn clauses in $R$.

There are a variety of inference relations defined in this paper. For any inference relation $\vdash$ and sets of ground formulas $\Sigma$ and $\Gamma$ we write $\Sigma \vdash \Gamma$ if $\Sigma \vdash \Psi$ for each $\Psi$ in $\Gamma$.

The inference relation $\vdash_R$ can be given a more direct syntactic characterization. This syntactic characterization is more useful in determining locality. In the following definitions and lemma, $\Sigma$ is a set of ground atomic formulas, and $\phi$ is a single ground atomic formula.

**Definition 2.** A *derivation* of $\Phi$ from $\Sigma$ using rule set $R$ is a sequence of ground atomic formulas $\Psi_1, \Psi_2, \ldots, \Psi_n$ such that $\Psi_n$ is $\Phi$ and for each $\Psi_i$ there exists a Horn clause $\Theta_1 \wedge \ldots \wedge \Theta_k \Rightarrow \Psi'$ in $R$ and a ground substitution $\sigma$ such that $\sigma[\Psi']$ is $\Psi_i$ and each formula of the form $\sigma[\Theta_i]$ is either a member of $\Sigma$ or a formula appearing in earlier than $\Psi_i$ in the derivation.

**Lemma 1:** $\Sigma \vdash_R \Phi$ if and only if there exists a derivation of $\Phi$ from $\Sigma$ using the rule set $R$.

The following restricted inference relation plays an important role in the analysis of locality.

**Definition 3.** We write $\Sigma \vdash\!\!\!\gg_R \Phi$ if there exists a derivation of $\Phi$ from $\Sigma$ such that every term appearing in the derivation appears as a subexpression of $\Phi$ or as a subexpression of some formula in $\Sigma$.

**Lemma 2:** (Tractability Lemma) [McAllester, 1993] For any finite rule set $R$ the inference relation $\vdash\!\!\!\gg_R$ is polynomial-time decidable.

**Proof:** Let $n$ be the number of terms that appear as subexpressions of $\Phi$ or of a formula in $\Sigma$. If $Q$ is a predicate-symbol of $k$ arguments that appears in the inference rules $R$ then there are at most $n^k$ formulas of the form $Q(s_1, \ldots, s_k)$ such that $\Sigma \vdash\!\!\!\gg_R Q(s_1, \ldots, s_k)$. Since $R$ is finite there is some maximum arity $k$ over all the predicate symbols that appear in $R$. The total number of ground atomic formulas that can be derived under the restrictions in the definition of $\vdash\!\!\!\gg_R$ is then of order $n^k$. Given a particular set of derived ground atomic formulas, one can determine whether any additional ground atomic formula can be derived by checking whether each rule in $R$ has an instance whose antecedents are all in the currently derived formulas — for a rule with $k'$ variables, there are only $n^{k'}$ instances to check, and each instance can be checked in polynomial time. Thus, one can extend the set of derived formulas by checking polynomially many instances, each in polynomial time; and the set of derived formulas can only be extended at most polynomially many times. The lemma then follows. ❑

Clearly, if $\Sigma \vdash\!\!\!\gg_R \Phi$ then $\Sigma \vdash_R \Phi$. But the converse does not hold in general. By definition, if the converse holds then $R$ is local.

**Definition 4.** [McAllester, 1993]: The rule set $R$ is *local* if the restricted inference relation $\vdash\mskip-14mu{\scriptstyle\gg}_R$ is the same as the unrestricted relation $\vdash_R$.

Clearly, if $R$ is local then $\vdash_R$ is polynomial-time decidable.

## 3. CHARACTERIZING P WITH SUPERFICIAL RULES

In this section we consider predicates on first-order terms that are computable in polynomial time. The results stated require a somewhat careful definition of a polynomial-time predicate on first-order terms.

**Definition 5.** A *polynomial-time predicate on terms* is a predicate $\tilde{Q}$ on one or more first-order terms which can be computed in polynomial time in the sum of the DAG sizes of its arguments.

**Definition 6.** A rule set is *superficial* if any term that appears in the conclusion of a rule also appears in some antecedent of that rule.

**Theorem 1:** (Superficial Rule Set Representation Theorem) If $\tilde{Q}$ is a polynomial-time predicate on $k$ first-order terms of a fixed finite signature, then there exists a superficial rule set $R$ such that for any first-order terms $t_1, \ldots, t_n$ from this signature, we have that $\tilde{Q}$ is true on arguments $t_1, \ldots, t_k$ if and only if $\texttt{INPUT}(t_1, \ldots, t_k) \vdash_R \texttt{ACCEPT}$.

As an example consider the "Acyclic" predicate on directed graphs — the predicate that is true of a directed graph if and only if that graph has no cycles. It is well-known that acyclicity is a polynomial-time property of directed graphs. This property has a simple definition using superficial rules with one level of stratification — if a graph is not cyclic then it is acyclic. The above theorem implies that the acyclicity predicate can be defined by superficial rules without any stratification. The unstratified rule set for acyclicity is somewhat complex and rather than give it here we give a proof of the above general theorem. The proof is rather technical, and casual readers are advised to skip to the next section.

**Proof:** (Theorem 1) We only consider predicates of one argument. The proof for predicates of higher arity is similar. Let $\tilde{Q}$ be a one argument polynomial-time computable predicate on terms, i.e., a predicate on terms such that one can determine in polynomial time in the DAG size of a term $t$ whether or not $\tilde{Q}(t)$ holds. Our general approach is to construct a database from $t$ such that the property $\tilde{Q}$ of terms can be viewed as a polynomial-time computable property of the database (since the term $t$ can be extracted from the database and then $\tilde{Q}(t)$ computed. We can then get a datalog program for computing this property of the database, given a total ordering of the database individuals, using the result of Immerman and Vardi [Immerman, 1986], [Vardi, 1982]. The proof finishes by showing how superficial rules can be given that construct the required database from $t$ and the required ordering of the database individuals. The desired superficial rule set is then the combination of the datalog program and the added rules for constructing the database and the ordering. We now argue this approach in more detail.

We first describe the database $\Sigma_t$ that will represent the term $t$. For each subterm $s$ of $t$ we introduce a database individual $c_s$, i.e., a new constant symbol unique to the term $s$. We have assumed that the predicate $\tilde{Q}$ is defined on terms constructed from a fixed finite signature, i.e., a fixed finite set of constant and function symbols. We will consider constants to be functions of no arguments. For each function symbol $f$ of $n$ arguments in this finite signature we introduce a database relation $P_f$ of $n+1$ arguments, i.e., $P_f$ is a $n+1$-ary predicate symbol. Now for any term $t$ we define $\Sigma_t$ to be the set of ground formulas of the form $P_f(c_{f(s_1, \ldots, s_n)}, c_{s_1}, \ldots, c_{s_n})$ where $f(s_1,\ldots,s_n)$ is a subterm of $t$ (possibly equal to $t$). The set $\Sigma_t$ should be viewed as a database with individuals $c_s$ and relations $P_f$. Let $\Gamma$ be a set of formulas of the form $\mathtt{S}(c_s, c_u)$ where $s$ and $u$ are subterms of $t$ such that $\mathtt{S}$ represents a successor relation on the individuals of $\Sigma_t$, i.e., there exists a bijection $\rho$ from the individuals of $\Sigma_t$ to consecutive integers such that $\mathtt{S}(s, u)$ is in $\Gamma$ if and only if $\rho(u) = \rho(s) + 1$. The result of Immerman and Vardi [Immerman, 1986], [Vardi, 1982] implies that for any polynomial-time property of the set $\Sigma_t \cup \Gamma$ there exists a datalog program $R$ such that $\Sigma_t \cup \Gamma$ has the given property if and only if $\Sigma_t \cup \Gamma \vdash_R \mathtt{ACCEPT}$. Since the term $t$ can be easily recovered from the set $\Sigma_t$, $\tilde{Q}$ can be viewed as a polynomial-time property of $\Sigma_t$, and so there must exist a datalog program $R$ such that $\Sigma_t \cup \Gamma \vdash_R \mathtt{ACCEPT}$ if and only if $\tilde{Q}(t)$. We can assume without loss of generality that no rule in $R$ can derive new formulas involving the database predicates $P_f$. If $R$ has such rules they can be eliminated by introducing duplicate predicates $P_f'$, adding rules that copy $P_f$ facts to $P_f'$ facts, and then replacing $P_f$ by $P_f'$ in all the rules.

We now add to the rule set $R$ superficial rules that construct the formulas needed in $\Sigma_t$ and $\Gamma$ — these rules use a number of "auxiliary" relation symbols in their computations; we assume the names of these relation symbols are chosen after the choice of $R$ so that there are no occurrences of these relation symbols in $R$. First we define a "mentioned" predicate $\mathtt{M}$ such that $\mathtt{M}(s)$ is provable if and only if $s$ is a subterm of $t$.

$$\mathtt{INPUT}(t) \Rightarrow \mathtt{M}(t)$$
$$\mathtt{M}(f(x_1, \ldots, x_n)) \Rightarrow \mathtt{M}(x_i) \tag{6}$$

The second rule is a schema for all rules of this form where $f$ is one of the finite number of function symbols in the signature and $x_i$ is one of the variables $x_1$, …, $x_n$. Now we give rules (again via a schema) that construct the formula set $\Sigma_t$.

$$\mathtt{M}(f(x_1, \ldots, x_n)) \Rightarrow P_f(f(x_1, \ldots, x_n), x_1, \ldots, x_n) \tag{7}$$

Now we write a collection of rules to construct the formula set Γ, i.e., rules that define a successor relation on the terms in $t$. We start by defining a simple subterm predicate Su such that Su($u$, $v$) is provable if $u$ and $v$ are subterms of $t$ such that $u$ is a subterm of $v$. The second rule is again a schema for all rules of this form within the finite signature.

$$\text{M}(x) \Rightarrow \text{Su}(x, x)$$
$$\text{M}(f(x_1, \ldots, x_n)), \text{Su}(y, x_i) \Rightarrow \text{Su}(y, f(x_1, \ldots, x_n)) \tag{8}$$

We also need the negation of the subterm predicate, which we will call NI for "not in". To define this predicate we first need to define a "not equal" predicate NE such that NE($u$, $v$) is provable if and only if $u$ and $v$ are distinct subterms of the input $t$.

$$\text{M}(f(x_1, \ldots, x_n)), \text{M}(g(y_1, \ldots, y_m)) \Rightarrow$$
$$\text{NE}(f(x_1, \ldots, x_n), g(y_1, \ldots, y_m))$$
$$\text{M}(f(x_1, \ldots, x_i, \ldots, x_n)), \text{M}(f(x_1, \ldots, y_i, \ldots, x_n)), \text{NE}(x_i, y_i) \Rightarrow$$
$$\text{NE}(f(x_1, \ldots, x_i, \ldots, x_n), f(x_1, \ldots, y_i, \ldots, x_n)) \tag{9}$$

Instances of the first rule schema must have $f$ and $g$ distinct function symbols and in the second rule schema $x_i$ and $y_i$ occur at the same argument position and all other arguments to $f$ are the same in both terms. Now we can define the "not in" predicate NI such that NI($s$, $u$) if $s$ is not a subterm of $u$. We only give the rules for constants and functions of two arguments. The rules for functions of other numbers of arguments are similar. Instances of the first rule schema must have $c$ a constant symbol.

$$\text{NE}(x, c) \Rightarrow \text{NI}(x, c)$$
$$\text{NE}(z, f(x, y)), \text{NI}(z, x), \text{NI}(z, y) \Rightarrow \text{NI}(z, f(x, y)) \tag{10}$$

Now for any subterm $s$ of the input we simultaneously define a three-place "walk" relation W($s$, $u$, $w$) and a binary "last" relation L($s$, $u$). W($s$, $u$, $w$) will be provable if $s$ and $u$ are subterms of $w$ and $u$ is the successor of $s$ in a left-to-right preorder traversal of the subterms of $w$ with elimination of later duplicates. L($s$, $u$) will be provable if $s$ is the last term of the left-to-right preorder traversal of the subterms of $u$, again with elimination of later duplicates. In these definitions, we also use the auxiliary three-place relation W′($s$, $u$, $v$), where W′($s$, $u$, $f(w, v)$) means roughly that $s$ and $u$ are subterms of $v$ such that $u$ comes after $s$ in the preorder traversal of $v$ and every term between $s$ and $u$ in this traversal is a subterm of $w$. More precisely, W′($s$, $u$, $v$) is inferred if and only if $v$ has the form $f(x,y)$ such that there are occurrences of $s$ and $u$ in the pre-order traversal of $y$ (removing duplicates within $y$) where the occurrence of $u$ is later than the occurrence of $s$ and all terms in between these occurrences in the traversal are sub-terms of $x$. Using `W'` and `NI` together (see two different rules below) enables the construction of a preorder traversal of $y$ with subterms of $x$ removed that can be used to construct a preorder traversal of $f(x,y)$ with duplicates removed.

$$\Rightarrow \text{L}(c, c)$$
$$\text{M}(f(x, y)), \text{L}(ylast, y), \text{NI}(ylast, x) \Rightarrow \text{L}(ylast, f(x, y))$$
$$\text{M}(f(x, y)), \text{Su}(y, x), \text{L}(xlast, x) \Rightarrow \text{L}(xlast, f(x, y)) \quad (11)$$
$$\text{L}(ylast, y), \text{Su}(ylast, x), \text{NI}(y, x),$$
$$\text{W'}(flast, ylast, f(x, y)), \text{NI}(flast, x) \Rightarrow \text{L}(flast, f(x, y))$$

$$\text{M}(f(x, y)) \Rightarrow \text{W}(f(x, y), x, f(x, y))$$
$$\text{M}(f(x, y)), \text{W}(u, v, x) \Rightarrow \text{W}(u, v, f(x, y))$$
$$\text{M}(f(x, y)), \text{NI}(y, x), \text{L}(s, x) \Rightarrow \text{W}(s, y, f(x, y)) \quad (12)$$
$$\text{M}(f(x, y)), \text{W}(u, v, y) \Rightarrow \text{W'}(u, v, f(x, y))$$
$$\text{W'}(u, v, f(x, y)), \text{NI}(u, x), \text{NI}(v, x) \Rightarrow \text{W}(u, v, f(x, y))$$
$$\text{W'}(u, v, f(x, y)), \text{W'}(v, w, f(x, y)), \text{Su}(v, x) \Rightarrow \text{W'}(u, w, f(x, y))$$

Finally we define the successor predicate `S` in terms of `W`, as follows.

$$\text{INPUT}(z), \text{W}(x, y, z) \Rightarrow \text{S}(x, y) \quad (13)$$

Let $R'$ be the datalog program $R$ plus all of the above superficial rules. We now have that $\Sigma_t \cup \Gamma \vdash_R$ `ACCEPT` if and only if `INPUT`$(t) \vdash_{R'}$ `ACCEPT`, and the proof is complete. ❏ (Theorem 1)

## 4. CHARACTERIZING P WITH LOCAL RULES

Using the theorem of the previous section one can provide a somewhat different characterization of the complexity class $P$ in terms of local rule sets.

**Definition 7.** A rule set $R$ is *local* if for any set of ground atomic formulas $\Sigma$ and any single ground atomic formula $\Phi$, we have $\Sigma \vdash_R \Phi$ if and only if $\Sigma \vdash\!\!\!\!\!»_R \Phi$. We note that the tractability lemma (Lemma 2) implies immediately that If $R$ is local then $\vdash_R$ is polynomial-time decidable.

**Theorem 2:** (Local Rule Set Representation Theorem) If $\tilde{Q}$ is a polynomial-time predicate on first-order terms then there exists a local rule set $R$ such that for any first-order terms $t_1, \ldots, t_k$, we have that $\tilde{Q}$ is true on arguments $t_1, \ldots, t_k$ if and only if $\vdash_R Q(t_1, \ldots, t_k)$ where $Q$ is a predicate symbol representing $\tilde{Q}$.

Before giving a proof of this theorem we give a simple example of a local rule set for a polynomial-time problem. Any context-free language can be recognized in cubic time. This fact is easily proven by giving a translation of grammars into local rule sets. We represent a string of words using a constant symbol for each word and the binary function CONS to construct terms that represent lists of words. For each nonterminal symbol $A$ of the grammar we introduce a predicate symbol $P_A$ of two arguments where $P_A(x, y)$ will indicate that $x$ and $y$ are strings of words and that $y$ is the result of removing a prefix of $x$ that parses as category $A$. For each grammar production $A \rightarrow c$ where $c$ is a terminal symbol we construct a rule with no antecedents and the conclusion $P_A(\text{CONS}(c, x), x)$. For each grammar production $A \rightarrow B\ C$ we have the following inference rule:

$$P_B(x, y) \wedge P_C(y, z) \Rightarrow P_A(x, z). \tag{14}$$

Finally, we let $P$ be a monadic predicate which is true of strings generated by the distinguished start nonterminal $S$ of the grammar and add the following rule:

$$P_S(x, \text{NIL}) \Rightarrow P(x). \tag{15}$$

Let $R$ be this set of inference rules. $R$ is a local rule set, although the proof of locality is not entirely trivial. The rule set $R$ also has the property that $\vdash_R P(x)$ if and only if $x$ is a string in the language generated by the given grammar. General methods for analyzing the order of running time of local rule sets can be used to immediately give that these clauses can be run to completion in order $n^3$ time where $n$ is the length of the input string.[3] We have implemented a compiler for converting local rule sets to efficient inference procedures. This compiler can be used to automatically generate a polynomial-time parser from the above inference rules.

**Proof:** (Theorem 2) We now prove the above theorem for local inference relations from the preceding theorem for superficial rule sets. By the superficial rule-set representation theorem there must exist a superficial rule set $R$ such that for any first order terms $t_1, \ldots, t_k$ we have that $Q(t_1, \ldots, t_k)$ if and only if $\text{INPUT}(t_1, \ldots, t_k) \vdash_R \text{ACCEPT}$ where INPUT is a predicate symbol and ACCEPT is a distinguished proposition symbol. For each predicate symbol $S$ of $m$ arguments appearing in $R$ let $S'$ be a new predicate symbol of $k+m$ arguments. We define the rule set $R'$ to be the rule set containing the following clauses.

---

3. An analysis of the order of running time for decision procedures for local inference relations is given in [McAllester, 1993].

$$\Rightarrow \texttt{Input}'(x_1, \ldots, x_k, x_1, \ldots, x_k)$$

$$\texttt{Accept}'(x_1, \ldots, x_k) \Rightarrow Q(x_1, \ldots, x_k)$$

$$S'_1(x_1, \ldots, x_k, t_{1,1}, \ldots, t_{1,m_1}) \wedge \ldots \qquad (16)$$
$$\ldots \wedge S'_n(x_1, \ldots, x_k, t_{n,1}, \ldots, t_{n,m_n}) \Rightarrow S'(x_1, \ldots, x_k, s_1, \ldots, s_j)$$

where $S_1(t_{1,1}, \ldots, t_{1,m_1}) \wedge \ldots \wedge S_n(t_{n,1}, \ldots, t_{n,m_n}) \Rightarrow S(s_1, \ldots, s_j)$ is in $R$

Given the above definition we can easily show that $\vdash_R S'(t_1, \ldots, t_k, s_1, \ldots, s_m)$ if and only if $\texttt{Input}(t_1, \ldots, t_k) \vdash_R S(s_1, \ldots, s_m)$. Therefore, it follows that $\texttt{Input}(t_1, \ldots, t_k) \vdash_R \texttt{Accept}$ if and only if $\vdash_{R'} Q(t_1, \ldots, t_k)$. It remains only to show that $R'$ is local. Suppose that $\Sigma \vdash_{R'} \Phi$. We must show that $\Sigma \vdash\!\!\gg_{R'} \Phi$. Let $t_1, \ldots, t_k$ be the first $k$ arguments in $\Phi$. If $\Phi$ is $Q(t_1, \ldots, t_k)$ then either $\Phi$ is in $\Sigma$ (in which case the result is trivial), or we must also have $\Sigma \vdash_{R'} \texttt{Accept}'(t_1, \ldots, t_k)$ so that it suffices to prove the result assuming that $\Phi$ is the application of the primed version of a predicate appearing in $R$. Every derivation based on $R'$ involves formulas which all have the same first $k$ arguments — in particular, given that $\Sigma \vdash_{R'} \Phi$ we must have that $\Sigma' \vdash_{R'} \Phi$ where $\Sigma'$ is the set of formulas in $\Sigma$ that have $t_1, \ldots, t_k$ as their first $k$ arguments. Let $\Sigma''$ and $\Phi'$ be the result of replacing each formula $S'(t_1, \ldots, t_k, s_1, \ldots, s_m)$ by $S(s_1, \ldots, s_m)$ in $\Sigma'$ and $\Phi$, respectively. Since $\Sigma' \vdash_{R'} \Phi$ we must have $\texttt{Input}(t_1, \ldots, t_k) \cup \Sigma'' \vdash_R \Phi'$. But since $R$ is superficial every term in the derivation underlying $\texttt{Input}(t_1, \ldots, t_k) \cup \Sigma'' \vdash_R \Phi'$ either appears in some $t_i$ or appears in $\Sigma''$. This implies that every term in the derivation appears in either $\Sigma'$ or $\Phi$, and thus that $\Sigma \vdash\!\!\gg_{R'} \Phi$. ☐ (Theorem 2)

## 5. ANOTHER CHARACTERIZATION OF LOCALITY

In this section we give an alternate characterization of locality. This characterization of locality plays an important role in both the definition of bounded-local rule sets given in [McAllester, 1993] and in the notion of inductively-local rule sets given in the next section.

**Definition 8.** A *bounding set* is a set Y of ground terms such that every subterm of a member of Y is also a member of Y.

**Definition 9.** A ground atomic formula $\Psi$ is called a *label formula* of a bounding set Y if every term in $\Psi$ is a member of Y.

**Definition 10.** For any bounding set Y, we define the inference relation $\vdash\!\!\gg_{R,Y}$ to be such that $\Sigma \vdash\!\!\gg_{R,Y} \Phi$ if and only if there exists a derivation of $\Phi$ from $\Sigma$ such that every formula in the derivation is a label formula of the term set Y.

We have that $\Sigma \mapsto_R \Phi$ if and only if $\Sigma \mapsto_{R,Y} \Phi$ where Y is the set of all terms appearing as subexpressions of $\Phi$ or of formulas in $\Sigma$. The inference relation $\mapsto_{R,Y}$ can be used to give another characterization of locality. Suppose that $R$ is not local. In this case there must exist some $\Sigma$ and $\Phi$ such that $\Sigma \not\mapsto_R \Phi$ but $\Sigma \vdash_R \Phi$. Let Y be the set of terms that appear in $\Sigma$ and $\Phi$. We must have $\Sigma \not\mapsto_{R,Y} \Phi$. However, since $\Sigma \vdash_R \Phi$ we must have $\Sigma \mapsto_{R,Y'} \Phi$ for some finite superset Y' of Y. Consider "growing" the bounding set one term at a time, starting with the terms that appear in $\Sigma$ and $\Phi$.

**Definition 11.** A *one-step extension* of a bounding set Y is a ground term $\alpha$ that is not in Y but such that every proper subterm of $\alpha$ is a member of Y.

**Definition 12.** A feedback event for $R$ consists of a finite set $\Sigma$ of ground formulas, a ground formula $\Phi$, a bounding set Y containing all terms that appear in $\Sigma$ and $\Phi$, and a one-step extension $\alpha$ of Y such that $\Sigma \mapsto_{R,Y \cup \{\alpha\}} \Phi$, but $\Sigma \not\mapsto_{R,Y} \Phi$.

By abuse of notation, a feedback event will be written as $\Sigma \mapsto_{R,Y \cup \{\alpha\}} \Phi$.

**Lemma 3:** [McAllester, 1993]: $R$ is local if and only if there are no feedback events for $R$.

**Proof:** First note that if $R$ has a feedback event then $R$ is not local — if $\Sigma \mapsto_{R,Y \cup \{\alpha\}} \Phi$ then $\Sigma \vdash_R \Phi$ but if $\Sigma \not\mapsto_{R,Y} \Phi$ then $\Sigma \not\mapsto_R \Phi$. Conversely suppose that $R$ is not local. In that case there is some $\Sigma$ and $\Phi$ such that $\Sigma \not\mapsto_R \Phi$ but $\Sigma \mapsto_{R,Y} \Phi$ for some finite Y. By considering a least such Y one can show that a feedback event exists for $R$. ❑

The concepts of bounded locality and inductive locality both involve the concept of a feedback event. We can define bounded locality by first defining $C_R(\Sigma, Y)$ to be the set of formulas $\Psi$ such that $\Sigma \mapsto_{R,Y} \Psi$. $R$ is bounded-local if it is local and there exists a natural number $k$ such that whenever $\Sigma \mapsto_{R,Y \cup \{\alpha\}} \Psi$ there exists a $k$-step or shorter derivation of $\Psi$ from $C_R(\Sigma, Y)$ such that every term in the derivation is a member of $Y \cup \{\alpha\}$. As mentioned above, the set of the four basic inference rules for equality is bounded-local and there exists a procedure which can recognize the locality of any bounded-local rule set [McAllester, 1993]. The definition of inductive locality is somewhat more involved and is given in the next section.

## 6. INDUCTIVE LOCALITY

To define inductive locality we first define the notion of a feedback template. A feedback template represents a set of potential feedback events. We also define a backward chaining process which generates feedback templates from a rule set $R$.

We show that if there exists a feedback event for *R* then such an event will be found by this backchaining process. Furthermore, we define an "inductive" termination condition on the backchaining process and show that if the backchaining process achieves inductive termination then *R* is local.

Throughout this section we let *R* be a fixed but arbitrary set of Horn clauses. The inference relation $\mapsto\!\!\!\!\!\!\gg_{R,Y}$ will be written as $\mapsto\!\!\!\!\!\!\gg_Y$ with the understanding that *R* is an implicit parameter of the relation.

We define feedback templates as ground objects — they contain only ground first-order terms and formulas. The process for generating feedback templates is defined as a ground process — it only deals with ground instances of clauses in *R*. The ground process can be "lifted" using a lifting transformation. Since lifting is largely mechanical for arbitrary ground procedures [McAllester and Siskind, 1991], the lifting operation is only discussed very briefly here.

**Definition 13.** A *feedback template* consists of a set of ground atomic formulas $\Sigma$, a multiset of ground atomic formulas $\Gamma$, a ground atomic formula $\Phi$, a bounding set Y, and a one-step extension $\alpha$ of Y such that $\Phi$ and every formula in $\Sigma$ is a label formula of Y, every formula in $\Gamma$ is a label formula of $Y \cup \{\alpha\}$ that contains $\alpha$, and such that $\Sigma \cup \Gamma \mapsto\!\!\!\!\!\!\gg_{Y \cup \{\alpha\}} \Phi$.

By abuse of notation a feedback template will be written as $\Sigma, \Gamma \mapsto\!\!\!\!\!\!\gg_{Y \cup \{\alpha\}} \Phi$. $\Gamma$ is a multiset of ground atomic formulas, each of which is a label formula of $Y \cup \{\alpha\}$ containing $\alpha$, and such that the union of $\Sigma$ and $\Gamma$ allow the derivation of $\Phi$ relative to the bounding set $Y \cup \{\alpha\}$. A feedback template is a potential feedback event in the sense that an extension of $\Sigma$ that allows a derivation of the formulas in $\Gamma$ may result in a feedback event. The requirement that $\Gamma$ be a multiset is needed for the template-based induction lemma given below. Feedback templates for *R* can be constructed by backward chaining.

**Procedure for Generating a Template for *R*:**

1. Let $\Psi_1 \wedge \ldots \wedge \Psi_n \Rightarrow \Phi$ be a ground instance of a clause in *R*.

2. Let $\alpha$ be a term that appears in the clause but does not appear in the conclusion $\Phi$ and does not appear as a proper subterm of any other term in the clause.

3. Let Y be a bounding set that does not contain $\alpha$ but does contain every term in the clause other than $\alpha$.

4. Let $\Sigma$ be the set of antecedents $\Psi_i$ which do not contain $\alpha$.

5. Let $\Gamma$ be the set of antecedents $\Psi_i$ which do contain $\alpha$.

6. Return the feedback template $\Sigma, \Gamma \mapsto\!\!\!\!\!\!\gg_{Y \cup \{\alpha\}} \Phi$.

We let $T_0[R]$ be the set of all feedback templates that can be derived from $R$ by an application of the above procedure. We leave it to the reader to verify that $T_0[R]$ is a set of feedback templates. Now consider a feedback template $\Sigma, \Gamma \mapsto\!\!\!\gg_{Y \cup \{\alpha\}} \Phi$. We can construct a new template by backward chaining from $\Sigma, \Gamma \mapsto\!\!\!\gg_{Y \cup \{\alpha\}} \Phi$ using the following procedure.

**Procedure for Backchaining from $\Sigma, \Gamma \mapsto\!\!\!\gg_{Y \cup \{\alpha\}} \Phi$**

1. Let $\Theta$ be a member of $\Gamma$

2. Let $\Psi_1 \wedge \ldots \wedge \Psi_n \Rightarrow \Theta$ be a ground instance of a clause in $R$ that has $\Theta$ as its conclusion and such that each $\Psi_i$ is a label formula of $Y \cup \{\alpha\}$.

3. Let $\Sigma'$ be $\Sigma$ plus all antecedents $\Psi_i$ that do not contain $\alpha$.

4. Let $\Gamma'$ be $\Gamma$ minus $\Theta$ plus all antecedents $\Psi_i$ that contain $\alpha$.

5. Return the template $\Sigma', \Gamma' \mapsto\!\!\!\gg_{Y \cup \{\alpha\}} \Phi$.

In step 4 of the above procedure, $\Gamma'$ is constructed using multiset operations. For example, if the multiset $\Gamma$ contains two occurrences of $\Theta$, then "$\Gamma$ minus $\Theta$" contains one occurrence of $\Theta$. We need $\Gamma$ to be a multiset in order to guarantee that certain backchaining operations commute in the proof of the induction lemma below — in particular, we will use the fact that if a sequence of backchaining operations remove an element $\Theta$ of $\Gamma$ at some point, then there exists a permutation of that sequence of backchaining operations producing the same resulting template, but that removes $\Theta$ first.

For any set $T$ of feedback templates we define $B[T]$ to be $T$ plus all templates that can be derived from an element of $T$ by an application of the above backchaining procedure. It is important to keep in mind that by definition $B[T]$ contains $T$. We let $B^n[T]$ be $B[B[\cdots B[T]]]$ with $n$ applications of $B$.

**Definition 14.** A feedback template is called *critical* if $\Gamma$ is empty.

If $\Sigma, \emptyset \mapsto\!\!\!\gg_{Y \cup \{\alpha\}} \Phi$ is a critical template then $\Sigma \mapsto\!\!\!\gg_{Y \cup \{\alpha\}} \Phi$. If $\Sigma \not\mapsto\!\!\!\gg_{Y} \Phi$ then $\Sigma \mapsto\!\!\!\gg_{Y \cup \{\alpha\}} \Phi$ is a feedback event. By abuse of notation, a critical template $\Sigma, \emptyset \mapsto\!\!\!\gg_{Y \cup \{\alpha\}} \Phi$ such that $\Sigma \not\mapsto\!\!\!\gg_{Y} \Phi$ will itself be called a feedback event. The following lemma provides the motivation for the definition of a feedback template and the backchaining process.

**Lemma 4:** There exists a feedback event for $R$ if and only if there exists a $j$ such that $B^j[T_0[R]]$ contains a feedback event.

**Proof:** To prove the Lemma 4 suppose that there exists a feedback event for $R$. Let $\Sigma \mapsto\!\!\!\gg_{Y \cup \{\alpha\}} \Phi$ be a minimal feedback event for $R$, i.e., a feedback event for $R$ which minimizes the length of the derivation of $\Phi$ from $\Sigma$ under the bounding set $Y \cup \{\alpha\}$. The fact that this feedback event is minimal implies that every formula in the derivation other than $\Phi$ contains $\alpha$. To see this suppose that $\Theta$ is a formula in the derivation other than $\Phi$ that does not involve $\alpha$. We then have $\Sigma \mathrel{\vert\!\gg}_{Y \cup \{\alpha\}} \Theta$ and $\Sigma \cup \{\Theta\} \mathrel{\vert\!\gg}_{Y \cup \{\alpha\}} \Phi$. One of these two must be a feedback event — otherwise we would have $\Sigma \mathrel{\vert\!\gg}_Y \Phi$. But if one of these is a feedback event then it involves a smaller derivation than $\Sigma \mathrel{\vert\!\gg}_{Y \cup \{\alpha\}} \Phi$ and this contradicts the assumption that $\Sigma \mathrel{\vert\!\gg}_{Y \cup \{\alpha\}} \Phi$ is minimal. Since every formula other than $\Phi$ in the derivation underlying $\Sigma \mathrel{\vert\!\gg}_{Y \cup \{\alpha\}} \Phi$ contains $\alpha$, the template $\Sigma, \emptyset \mathrel{\vert\!\gg}_{Y \cup \{\alpha\}} \Phi$ can be derived by backchaining steps mirroring that derivation. ❑

The above lemma implies that if the rule set is not local then backchaining will uncover a feedback event. However, we are primarily interested in those cases where the rule set is local. If the backchaining process is to establish locality then we must find a termination condition which guarantees locality. Let $T$ be a set of feedback templates. In practice $T$ can be taken to be $B^j[T_0[R]]$ for some finite $j$. We define a "self-justification" property for sets of feedback templates and prove that if $T$ is self-justifying then there is no $n$ such that $B^n[T]$ contain a feedback event. In defining the self-justification property we treat each template in $T$ as an independent induction hypothesis. If each template can be "justified" using the set of templates as induction hypotheses, then the set $T$ is self-justifying.

**Definition 15.** We write $\Sigma, \Gamma \mathrel{\vert\!\gg}_{T,Y} \Phi$ if $T$ contains templates

$$
\begin{aligned}
\Sigma_1, \Gamma_1 &\mathrel{\vert\!\gg}_{Y \cup \{\alpha\}} \Psi_1 \\
\Sigma_2, \Gamma_2 &\mathrel{\vert\!\gg}_{Y \cup \{\alpha\}} \Psi_2 \\
&\ldots \\
\Sigma_k, \Gamma_k &\mathrel{\vert\!\gg}_{Y \cup \{\alpha\}} \Psi_k
\end{aligned}
\tag{17}
$$

where each $\Sigma_i$ is a subset of $\Sigma$, each $\Gamma_i$ is a subset of $\Gamma$ and $\Sigma \cup \{\Psi_1, \Psi_2, \ldots, \Psi_k\} \mathrel{\vert\!\gg}_Y \Phi$.

**Definition 16.** A set of templates $T$ is said to *justify* a template $\Sigma, \Gamma \mathrel{\vert\!\gg}_{Y \cup \{\alpha\}} \Phi$ if there exists a $\Theta \in \Gamma$ such that for each template $\Sigma', \Gamma' \mathrel{\vert\!\gg}_{Y \cup \{\alpha\}} \Phi$ generated by one step of backchaining from $\Sigma, \Gamma \mathrel{\vert\!\gg}_{Y \cup \{\alpha\}} \Phi$ by selecting $\Theta$ at step 1 of the backchaining procedure we have $\Sigma', \Gamma' \mathrel{\vert\!\gg}_{T,Y} \Phi$.

**Definition 17.** The set $T$ is called *self-justifying* if every member of $T$ is either critical or justified by $T$, and $T$ does not contain any feedback events.

**Theorem 3:** (Template-based Induction Theorem) If $T$ is self-justifying then no set of the form $B^n[T]$ contains a feedback event.

**Proof:** Consider a self-justifying set $T$ of templates. We must show that for every critical template $\Sigma, \emptyset \mathrel{\vert\!\gg}_{Y \cup \{\alpha\}} \Phi$ in $B^n[T]$ we have that $\Sigma \mathrel{\vert\!\gg}_Y \Phi$. The proof is by induction on $n$. Consider a critical template $\Sigma, \emptyset \models_{Y \cup \{\alpha\}} \Phi$ in $B^n[T]$ and assume the theorem for all critical templates in $B^j[T$ for $j$ less than $n$. The critical template $\Sigma, \emptyset \models_{Y \cup \{\alpha\}} \Phi$ must be derived by backchaining from some template $\Sigma', \Gamma' \models_{Y \cup \{\alpha\}} \Phi$ in $T$. Note that $\Sigma'$ must be a subset of $\Sigma$. If $\Gamma'$ is empty then $\Sigma'$ equals $\Sigma$ and $\Sigma \models_Y \Phi$ because $T$ is self-justifying and thus cannot contain any feedback events. If $\Gamma'$ is not empty then, since $T$ is self-justifying, we can choose a $\Theta$ in $\Gamma'$ such that for each template $\Sigma'', \Gamma'' \models_{Y \cup \{\alpha\}} \Phi$ derived from $\Sigma', \Gamma' \models_{Y \cup \{\alpha\}} \Phi$ by a single step of backchaining on $\Theta$ we have $\Sigma'', \Gamma'' \models_{T,Y} \Phi$. We noted above that backchaining operations commute (to ensure this we took $\Gamma$ to be a multiset rather than a set). By the commutativity of backchaining steps there exists a backchaining sequence from $\Sigma', \Gamma' \models_{Y \cup \{\alpha\}} \Phi$ to $\Sigma, \emptyset \models_{Y \cup \{\alpha\}} \Phi$ such that the first step in that sequence is a backchaining step on $\Theta$. Let $\Sigma^*, \Gamma^* \models_{Y \cup \{\alpha\}} \Phi$ be the template that results from this first backchaining step from $\Sigma', \Gamma' \models_{Y \cup \{\alpha\}} \Phi$. Note that $\Sigma^*$ is a subset of $\Sigma$. We must now have $\Sigma^*, \Gamma^* \models_{T,Y} \Phi$. By definition, $T$ must contain templates

$$\Sigma_1, \Gamma_1 \models_{Y \cup \{\alpha\}} \Psi_1$$
$$\Sigma_2, \Gamma_2 \models_{Y \cup \{\alpha\}} \Psi_2 \qquad (18)$$
$$\ldots$$
$$\Sigma_k, \Gamma_k \models_{Y \cup \{\alpha\}} \Psi_k$$

such that each $\Sigma_i$ is a subset of $\Sigma^*$, each $\Gamma_i$ is a subset of $\Gamma^*$, and $\Sigma^* \cup \{\Psi_1, \Psi_2, \ldots, \Psi_k\} \models_Y \Phi$. Note that each $\Sigma_i$ is a subset of $\Sigma$. Since $\Gamma_i$ is a subset of $\Gamma^*$ there must be a sequence of *fewer than n* backchaining steps that leads from $\Sigma_i, \Gamma_i \models_{Y \cup \{\alpha\}} \Psi_i$ to a critical template $\Sigma'_i, \emptyset \models_{Y \cup \{\alpha\}} \Psi_i$ such that $\Sigma'_i$ is a subset of $\Sigma$. This critical template is a member of $B^j[T]$ for $j$ less than $n$ and so by our induction hypothesis this template cannot be a feedback event; as a consequence we have $\Sigma'_i \models_Y \Psi_i$ and thus $\Sigma \models_Y \Psi_i$. But if $\Sigma \models_Y \Psi_i$ for each $\Psi_i$, and $\Sigma \cup \{\Psi_1, \Psi_2, \ldots, \Psi_k\} \models_Y \Phi$, then $\Sigma \models_Y \Phi$. ☐ (Template-based Induction Theorem)

The following corollary then follows from Theorem 3 along with Lemmas 3 and 4:

**Corollary 1:** If $B^n[T_0[R]]$ is self-justifying, for some $n$, then $R$ is local.

We now come the main definition and theorem of this section.

**Definition 18.** A rule set $R$ is called *inductively-local* if there exists some $n$ such that $B^n[T_0[R]]$ is self-justifying.

**Theorem 4:** There exists a procedure which, given any finite set $R$ of Horn clauses, will terminate with a feedback event whenever $R$ is not local, terminate with "success" whenever $R$ is inductively-local, and fail to terminate in cases where $R$ is local but not inductively-local.

The procedure is derived by lifting the above ground procedure for computing $B^n[T_0[R]]$. Lifting can be formalized as a mechanical operation on arbitrary nondeterministic ground procedures [McAllester and Siskind, 1991]. In the lifted version the infinite set $B^j[T_0[R]]$ is represented by a finite set of "template schemas" each of which consists of a template expression $\Sigma, \Gamma \mathrel{\vdash\mkern-9mu\gg}_{Y \cup \{\alpha\}} \Phi$ involving variables plus a set of constraints on those variables. We have implemented the resulting lifted procedure and used it to verify the locality of a variety of rule sets, including the rule set given as equation (5) above for reasoning about lattices. This procedure is also useful for designing local rule sets — when applied to a nonlocal rule set the procedure returns a feedback event that can often be used to design additional rules that can be added to the rule set to give a local rule set computing the same inference relation.

## 7. LOCALITY IS UNDECIDABLE

We prove that locality is undecidable by reducing the Halting problem. Let $M$ be a specification of a Turing machine. We first show one can mechanically construct a local rule set $R$ with the property that the machine $M$ halts if and only if there exists a term $t$ such that $\vdash_R H(t)$ where $H$ is a monadic predicate symbol. Turing machine computations can be represented by first-order terms and the formula $H(t)$ intuitively states that $t$ is a term representing a halting computation of $M$.

To prove this preliminary result we first construct a superficial rule set $S$ such that $M$ halts if and only if there exists a term $t$ such that $\texttt{INPUT}(t) \vdash_S H(t)$. The mechanical construction of the superficial rule set $S$ from the Turing machine $M$ is fairly straightforward and is not given here. We convert this superficial rule set $S$ to a local rule set $R$ as follows. For each predicate symbol $Q$ of $m$ arguments appearing in $S$ let $Q'$ be a new predicate symbol of $m+1$ arguments. The rule set $R$ will be constructed so that $\vdash_R Q(t, s_1, ..., s_m)$ if and only if $\texttt{INPUT}(t) \vdash_R Q(t, s_1, ..., s_m)$. We define the rule set $R$ to be the rule set containing the following clauses: $\Rightarrow \texttt{Input}'(x, x)$, $H'(x, x) \Rightarrow H(x)$, and each clause of the form

$$Q'_1(x, t_{1,1}, ..., t_{1,m_1}) \wedge ... \qquad (19)$$
$$... \wedge Q'_n(x, t_{n,1}, ..., t_{n,m_n}) \Rightarrow W'(x, s_1, ..., s_j)$$

where $Q_1(t_{1,1}, ..., t_{1,m_1}) \wedge ... \wedge Q_n(t_{n,1}, ..., t_{n,m_n}) \Rightarrow W(s_1, ..., s_j)$ is in $R$. By the design of $R$ we can easily show that $\vdash_R Q'(t, s_1, ..., s_m)$ if and only if $\texttt{INPUT}(t) \vdash_R Q(t, s_1, ..., s_m)$, and so it directly follows that $\texttt{INPUT}(t) \vdash_S H(t)$ if and only if $\vdash_R H(t)$. So the Turing machine $M$ halts if and only if $\vdash_R H(t)$ for some term $t$, as desired. The proof that the rule set *R* is local closely follows the proof that *R'* is local in the Local Rule Set Representation Theorem proven above (Theorem 2).

We have now constructed a local rule set *R* with the property that *M* halts if and only if there exists some term *t* such that $\vdash_R H(t)$. Now let *R'* be *R* plus the single clause $H(x) \Rightarrow \text{HALTS}$ where HALTS is a new proposition symbol. We claim that *R'* is local if and only if *M* does not halt. First note that if *M* halts then we have both $\vdash_{R'} \text{HALTS}$ and $\not\mapsto_{R'} \text{HALTS}$ so *R* is not local. Conversely, suppose that *M* does not halt. In this case we must show that *R'* is local. Suppose that $\Sigma \vdash_{R'} \Phi$. We must show that $\Sigma \mapsto_{R'} \Phi$. Suppose $\Phi$ is some formula other than HALTS. In this case $\Sigma \vdash_{R'} \Phi$ is equivalent to $\Sigma \vdash_R \Phi$. Since *R* is local we must have $\Sigma \mapsto_R \Phi$ and thus $\Sigma \mapsto_{R'} \Phi$. Now suppose $\Phi$ is the formula HALTS. If HALTS is a member of $\Sigma$ then the result is trivial so we assume that HALTS is not in $\Sigma$. Since $\Sigma \vdash_{R'} \text{HALTS}$ we must have $\Sigma \vdash_{R'} H(c)$ for some term *c*. This implies that $\Sigma \vdash_R H(c)$ and thus $\Sigma \mapsto_R H(c)$ and $\Sigma \mapsto_{R'} H(c)$. To show $\Sigma \mapsto_{R'} \text{HALTS}$ it now suffices to show that *c* is mentioned in $\Sigma$. By the preceding argument we have $\Sigma \mapsto_R H(c)$. Since the rule set *R* was generated by the construction given above, we have that every inference based on a clause in *R* is such that every formula in the inference has the same first argument. This implies that $\Sigma' \mapsto_R H(c)$ where $\Sigma'$ is the set of all formulas in $\Sigma$ that have *c* as a first argument. We have assumed that *M* does not halt, and thus $\not\vdash_R H(c)$. Hence $\Sigma'$ must not be empty. Since every formula in $\Sigma'$ mentions *c*, and $\Sigma'$ is contained in $\Sigma$, we can conclude that $\Sigma$ must mention *c* — thus since $\Sigma \mapsto_R H(c)$ we have $\Sigma \mapsto_{R'} \text{HALTS}$.

## 8. OPEN PROBLEMS

In closing we note some open problems. There are many known examples of rule sets which are not local and yet the corresponding inference relation is polynomial-time decidable. In all such cases we have studied there exists a conservative extension of the rule set which is local. We conjecture that for every rule set *R* such that $\vdash_R$ is polynomial-time decidable there exists a local conservative extension of *R*. Our other problems are less precise. Can one find a "natural" rule set that is local but not inductively local? A related question is whether there are useful machine recognizable subclasses of the local rule sets other than the classes of bounded-local and inductively-local rule sets?

### Acknowledgements

We would like to thank Franz Baader for his invaluable input and discussions. Robert Givan was supported in part by National Science Foundation Award No. 9977981-IIS.